\documentstyle[osa,manuscript]{revtex}

\begin{document}
\title{\bf A Comment on "Semiquantum Chaos"}
\author{L. Faccioli, F. Finelli, G. P. Vacca, G. Venturi}

\address{Dept. of Physics and INFN, Bologna, Italy.}

\def\beq{\begin{equation}}
\def\eeq{\end{equation}}
\def\bea{\begin{eqnarray}}
\def\eea{\end{eqnarray}}
\def\noi{\noindent}
\maketitle

\begin{abstract}
The identification of the particle creation and distruction operators
is discussed.
\end{abstract}

\vskip 1.5truecm

In a letter \cite{cop1} Cooper et al. (henceforth CDMS)
considered a system, in which a classical oscillator interacts with a
purely quantum mechanical oscillator, described by the classical Lagrangian:
\beq
L= \frac{1}{2} \dot{x}^2 + \frac{1}{2} \dot{A}^2 -\frac{1}{2}(m^2+
e^2 A^2)x^2  
\label{lagra}
\eeq

The scope of this comment is to correctly identify the particle creation
and destruction operators. The corresponding time dependent occupation number
(differing from eq. (20) of CDMS) leads to changes in analytical and
numerical results.
Our results are obtained through the use of the Born-Oppenheimer 
approach (which is pertinent in the presence of two mass, or time, scales)
and the method of adiabatic invariants \cite{lewis}.

From (\ref{lagra}) one obtains for a state of energy $E$ a 
Schr\"odinger equation:
\beq
\frac{1}{2}\Bigl( -\hbar^2 \frac{\partial^2}{\partial x^2} + \omega^2 x^2
- \hbar^2 \frac{\partial^2}{\partial A^2} -2 E \Bigr) \Psi_E(x,A)=0
\label{schr}
\eeq
where $\omega^2=m^2+e^2 A^2$. On factorizing $\Psi_E(x,A)=\psi(A) \chi(x,A)$ 
(we omit the index $E$ for simplicity) one obtains the coupled equations:
\beq
\bigl[ -\frac{\hbar^2}{2} D^2 -E + \langle \hat{H_x} \rangle \bigr] \psi =
\frac{\hbar^2}{2} \langle \bar{D}^2 \rangle \psi 
\label{coupl1}
\eeq
\beq
\bigl( \hat{H_x}-\langle \hat{H_x} \rangle \bigr) \chi - \frac{\hbar^2}{\psi}
(D\psi) \bar{D} \chi = \frac{\hbar^2}{2}\bigl( \bar{D}^2-
\langle \bar{D}^2 \rangle \bigr) \chi
\label{coupl2}
\eeq
where (\ref{coupl2}) exists where $\psi$ has support; for an operator 
$\hat{O}$, $\langle \hat{O} \rangle=\int dx \  \chi^{*} \hat{O} \chi \ / 
\int dx \ \chi^{*}\chi$, $D=\frac{\partial}{\partial A} + \langle 
\frac{\partial}{\partial A} \rangle$, $\bar{D}= \frac{\partial}{\partial A} - 
\langle \frac{\partial}{\partial A} \rangle$ and $\hat{H}_x$ is given by the 
first two terms on the LHS of (\ref{schr}).
On considering the semiclassical approximation to 
$\psi \simeq \frac{1}{\sqrt{\dot{A}}} exp(-\int^A 
\langle \frac{\partial}{\partial A'} \rangle dA' + \frac{i}{\hbar} 
\int^A \dot{A'} dA')$
and neglecting the RHS of (\ref{coupl1}) (fluctuations), one reproduces the 
Hamilton-Jacobi equation for $A$ (eq. (13) of CDMS) and
consequently the time evolution equations.
The same semiclassical approximation for $\psi$ and the neglect of the RHS
in eq. (\ref{coupl2}) leads to the Schr\"odinger equation:
\beq
\bigl( \hat{H}_x - i \hbar \frac{\partial}{\partial t} \bigr) \chi_s = 0
\eeq
where $\chi_s=exp[- \int^t dt' (\frac{i}{\hbar} \langle \hat{H}_x \rangle 
 + \langle \frac{\partial}{\partial A'} \rangle  \dot{A}') ] \chi$. 
$\hat{H}_x$ can be factorized as:
\beq
\hat{H}_x= \hbar \omega (b^\dagger b +\frac{1}{2})
\label{hami}
\eeq
where 
$b=\sqrt{\frac{\omega}{2\hbar}}(x+\frac{\hbar}{\omega}
\frac{\partial}{\partial x})$ thus allowing us to identify 
the charged particle creation and destruction operators.

In order to obtain solutions to the Schr\"odinger equation it is convenient 
to introduce the Hermitian adiabatic invariant (satisfying 
$\frac{\partial \hat{I}}{\partial t}+\frac{1}{i \hbar} [\hat{I},\hat{H}]=0$):
\beq
\hat{I}=\hbar \bigl( a^\dagger a+\frac{1}{2}\bigr)
\label{invquad}
\eeq
where 
$a=e^{i \theta} \bigl( \frac{\Omega}{2\hbar}\bigr)^{\frac{1}{2}}
\bigl[ x \bigl(1+\frac{i}{2} \frac{\dot{\Omega}}{\Omega^2} \bigr) +
\frac{\hbar}{\Omega} \frac{\partial}{\partial x} \bigr]$
is a linear (non-hermitian) adiabatic invariant, which corresponds to the
same operator used by CDMS in the Schr\"odinger representation,
and  $\theta=\int^t dt' \ \Omega$.
Further $\Omega$ satisfies
\beq
\frac{1}{2} \frac{\ddot{\Omega}}{\Omega} -\frac{3}{4} \bigl(
\frac{\dot{\Omega}}{\Omega} \bigr)^2 + \Omega^2 = \omega^2
\label{pinney}
\eeq
On solving eq. (\ref{pinney}) 
one knows the evolution of the quantum system.
The $a$ and $b$ operators are related by a Bogolubov trasformation: 
however while our $a$'s agree with those of CDMS,
the $b$'s do not. Indeed one sees that with their choice 
the RHS of (\ref{hami}) corresponds to
\beq
-\frac{\hbar^2}{2}
\frac{\partial^2}{\partial x^2} +\frac{x^2}{2}
\bigl[ \omega^2+\frac{1}{4} \bigl(\frac{\dot{\omega}}{\omega} \bigr)^2 \bigl]
- i \hbar \frac{\dot{\omega}}{4 \omega} \{x,\frac{\partial}{\partial x} \}
\eeq
On defining the vacuum state 
by $a |0 \rangle = 0$ we can compute the following average quantities
for $| \chi \rangle = | 0 \rangle$:
\beq
\langle x^2 \rangle = \frac{\hbar}{2 \Omega} \quad ; \quad
\langle \dot{x}^2 \rangle = \frac{\hbar}{2} \bigl( \Omega + \frac{1}{4} 
\frac{\dot{\Omega} ^2 }{\Omega^3} \bigr)
\label{ave1}
\eeq
Finally the average time dependent particle number is:
\beq
\langle b^\dagger b \rangle = \frac{1}{4} \bigl( \frac{\omega}{\Omega}
+\frac{\Omega} {\omega} + \frac{1}{4} \frac {\dot{\Omega}^2} {\omega \Omega^3}
\bigr) - \frac{1}{2}
\label{num}
\eeq
It is straightforward to verify that the expressions in (\ref{ave1}) agree
with the corresponding ones of CDMS while (\ref{num})
does not.
In order to better quantify this we have approximately evaluated
eq. (\ref{num}) in the two cases (ours and ref \cite{cop1}) for
$e^2A^2/m^2 < 1$ and find that they differ by a term:
\beq
\frac{e^4}{16 m^6}\dot{A}^2 A^2 + O(e^6)
\eeq
which, depending on $A$, can be significant.

We have also seen, by a numerical analysis
similar to the one of CDMS, that although the
general time behaviour of $\langle b^\dagger b \rangle $ is similar,
our expression (\ref{num}) exhibits more structure (see figure).
Let us end by noting that one may also estimate the neglected terms,
finding that both the RHS of eq. (\ref{coupl1}) and (\ref{coupl2}) and
the non-leading term arising from the prefactor in the semiclassical
limit for $\psi$ give corrections of the order 
$\frac{\hbar^2 e^2}{m^2} \langle b^\dagger b \rangle$.

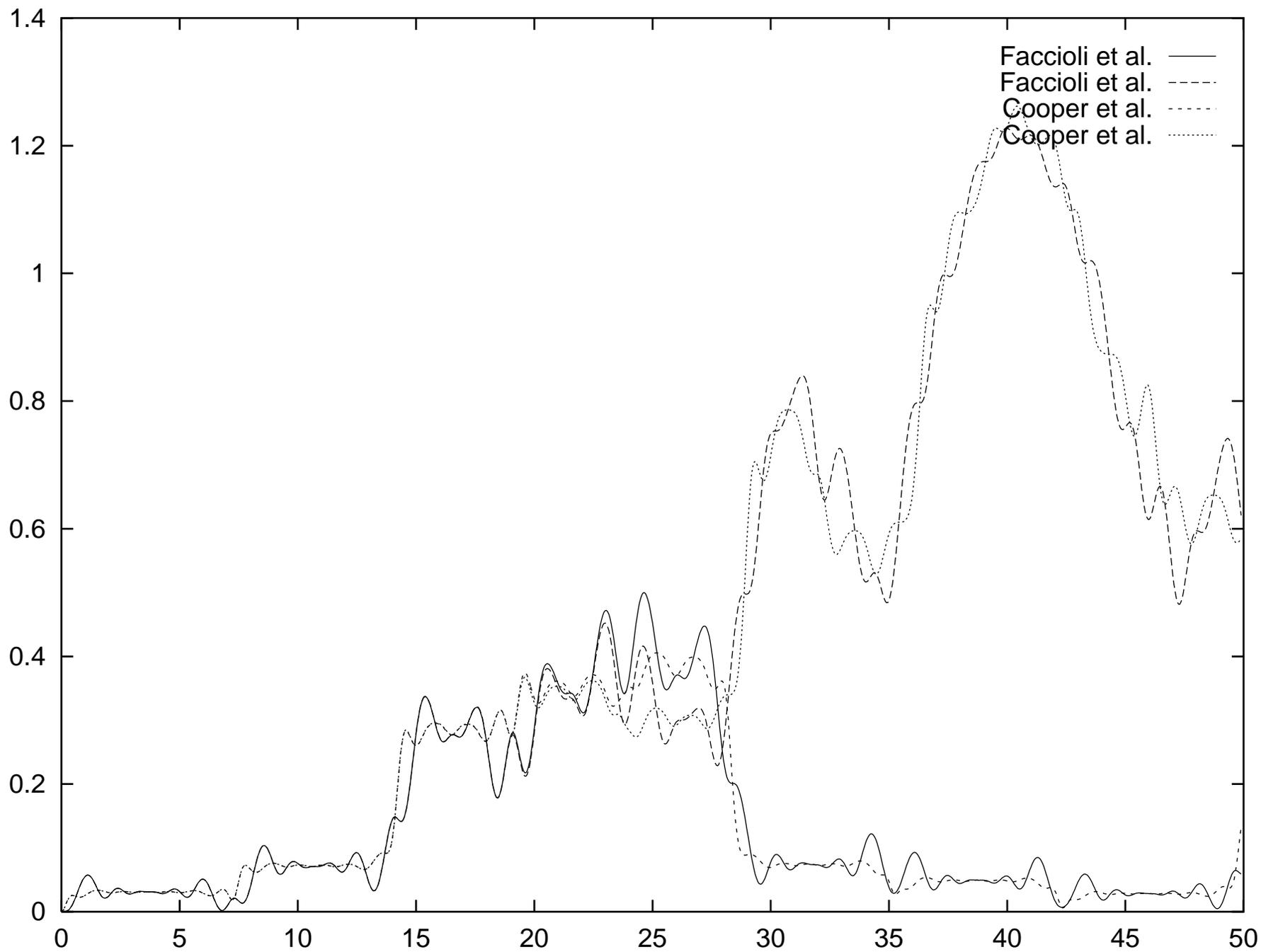
\end{document}